# Extracting Entities and Topics from News and Criminal Record

Quang Pham [a], Marija Stanojevic [b] and Zoran Obradovic [c]

*Center for Data Analytics and Biomedical Informatics, Computer and Information Sciences Department, Temple University*

*a) quang.tran.pham.temple.edu*
*b) marija.stanojevic@temple.edu*
*c) zoran.obradovic.temple.edu*

## INTRODUCTION

The goal of this paper is to summarize methodologies used in extracting entities and topics from a database of criminal records and from a database of newspapers. Statistical models had successfully been used in studying the topics of roughly 300,000 New York Times articles. In addition, these models had also been used to successfully analyze entities related to people, organizations, and places (D Newman, 2006). Additionally, analytical approaches, especially in hotspot mapping, were used in some researches with an aim to predict crime locations and circumstances in the future, and those approaches had been tested quite successfully (S Chainey, 2008). Based on the two above notions, this research was performed with the intention to apply data science techniques in analyzing a big amount of data, selecting valuable intelligence, clustering violations depending on their types of crime, and creating a crime graph that changes through time. In this research, the task was to download criminal datasets from Kaggle and a collection of news articles from Kaggle and EAGER project databases, and then to merge these datasets into one general dataset. The most important goal of this project was performing statistical and natural language processing methods to extract entities and topics as well as to group similar data points into correct clusters, in order to understand public data about U.S related crimes better.

## DATASETS DESCRIPTION

### A. Kaggle Datasets

Kaggle datasets contain a total of 10 databases. Five of them are databases of committed crime in specific cities and the other five contain a report on the cases of homicide in the U.S., a fatal police shootings record in the U.S., a record of mass shootings in the U.S, global terrorism database and a database with news articles related to criminal activities. Several datasets contained information about international crimes, but it was out of our scope of interest, so those data points were eliminated in the processing steps. In Table I, the number of rows measures number of total cases in each state, while the number of columns represents the variety of records for each violation. Some of the popular attributes that appeared in the five specific datasets are location, date, crime types, and details. However, features in different datasets have named in different ways even though they convey the same meaning.

**TABLE I.** Details of 5 states datasets.

| State | Number of rows | Number of columns |
|---|---|---|
| Boston[1] | 318,073 | 6 |
| Chicago[2] | 1,048,575 | 7 |
| Denver [3] | 460,919 | 6 |
| Philadelphia [4] | 1,048,675 | 5 |
| San Francisco[5] | 1,048,575 | 6 |

Table II demonstrates dataset size details of the 5 remaining datasets. In the column "Number of rows", the minimum value is 1681 and the maximum value is 638454 while in the next column, the minimum number of features is 10 and the maximum is 133, as shown in the column "Number of columns. This validation strongly proved that these datasets have diverse size and also provided explanations for some essential factors of each dataset. For example, the Homicide dataset has a very high number of samples because this dataset has been recorded for a longer time, and because homicide occurs more often than does terrorism or mass shootings (WA Pridemore, 2008). One more reason that leads to the high number of homicide cases is serial murder record, which means one penetrator committed more than one crime (RM Holmes, 1985). In another example, the number of recorded attributes in a single terrorism case is 133, which is very high compare to the numbers of other datasets. Since terrorism is a very serious and complex crime, a lot of attributes must be considered in a single case such as victims, convicts, groups, or purposes, which lead to the high dimension of the dataset (G LaFree, 2009).

One more detail to notice is that attributes in News datasets are totally different from attributes in nine other datasets. Moreover, attributes related to an article normally contain the time, source, headline, and content, while attributes in nine other datasets contain detailed information about a crime, the information that is used primarily for investigating and storing criminal records instead of for publishing.

---

[1] https://www.kaggle.com/ankkur13/boston-crime-data
[2] https://www.kaggle.com/chicago/chicago-crime
[3] https://www.kaggle.com/paultimothymooney/denver-crime-data
[4] https://www.kaggle.com/mchirico/philadelphiacrimedata
[5] https://www.kaggle.com/roshansharma/sanfranciso-crime-dataset

**Table II Statistics of databases**

| Database | Number of rows | Number of columns |
|---|---|---|
| News[6] | 142,570 | 10 |
| Police kills[7] | 2335 | 14 |
| Homicide[8] | 638,454 | 18 |
| Global Terrorism[9] | 2836 | 138 |
| US Mass Shooting[10] | 1681 | 13 |

## B. Articles database

A dataset of 1,048,575 articles, from 22,770 online news outlets had been downloaded as part of the EAGER news project database before the start of this project. This dataset contains news from period between October 2015 and January 2017. In this database, there are popular newspaper institutions such as NYT, WSJ or USNEWS, and some obscure sources than with less than number ten published articles in that period, such as "Allies is Wired" or "WirelessWatch.

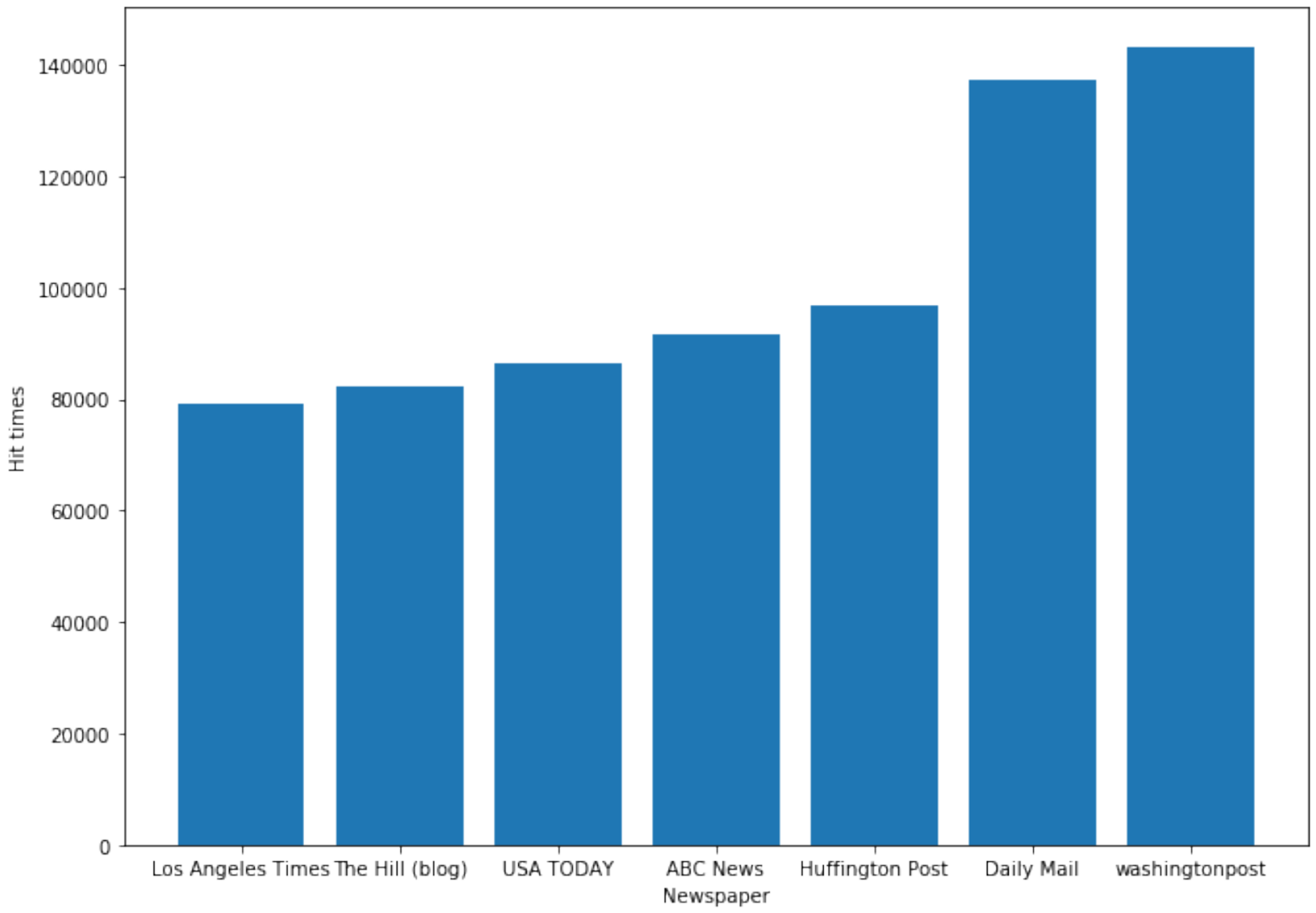

**FIGURE 1.** Most popular articles source in Eager.

Every news institutions in the dataset has the Hit Times value, which is the total number of articles published from each institution. A high Hit Times value can strongly support the reputation and the certitude of an institution, especially in political aspects (M Schudson, 2002).Figure 1 shows seven most popular sources have a very high value of Hit Times, such as the "washingtonpost"

6 https://www.kaggle.com/snapcrack/all-the-news
7 https://www.kaggle.com/kwullum/fatal-police-shootings-in-the-us
8 https://www.kaggle.com/murderaccountability/homicide-reports
9 https://www.kaggle.com/START-UMD/gtd
10 https://www.kaggle.com/zusmani/us-mass-shootings-last-50-years

with more than 140,000 articles, or "the Daily Mail" with nearly 139,000 articles. Additionally, the standard deviation, the average value, and percentile values of the Hit Times collection, which are shown in Table III, strongly support the big gap between popular and unpopular institutions. In Table III, the standard deviation value is 3526.024; the average value is 598; the median is 40; the maximum is 143,300. This evidence clearly shows a wide range of publications activity among outlets, and the median value equals to 40.0 strongly supports this claim. Since most of the popular news institutions are from the U.S, where the development of digital media can help institutions attract the attention of a diverse group of audiences by using multi-channel service, this famous institutions can have more resources to help them ensuring quantity and quality of their articles (N Thurman, 2014).

**Table III.** Statistics of articles sources

| | Value |
|---|---|
| Standard Deviation | 3526.024 |
| Mean | 598 |
| 25th percentile | 13.0 |
| 50th percentile | 40.0 |
| 75th percentile | 190.0 |
| 100th percentile | 143,300 |

The content of thesearticles sheds light on categories with the highest number of articles by counting the number of times each category ID appears. The most popular category is "Top Stories", with 46,186 articles, while the second and third are "World" with 2608 and "U.S." with 1923, respectively. Following the first three categories are business, politics, and sport. However, the gap between the first place and second place in popularity is very big, around 40,000 articles, suggesting that when an event is published, it will have a very high chance to be tagged as "Top Stories", based on the priority of the uploading time.

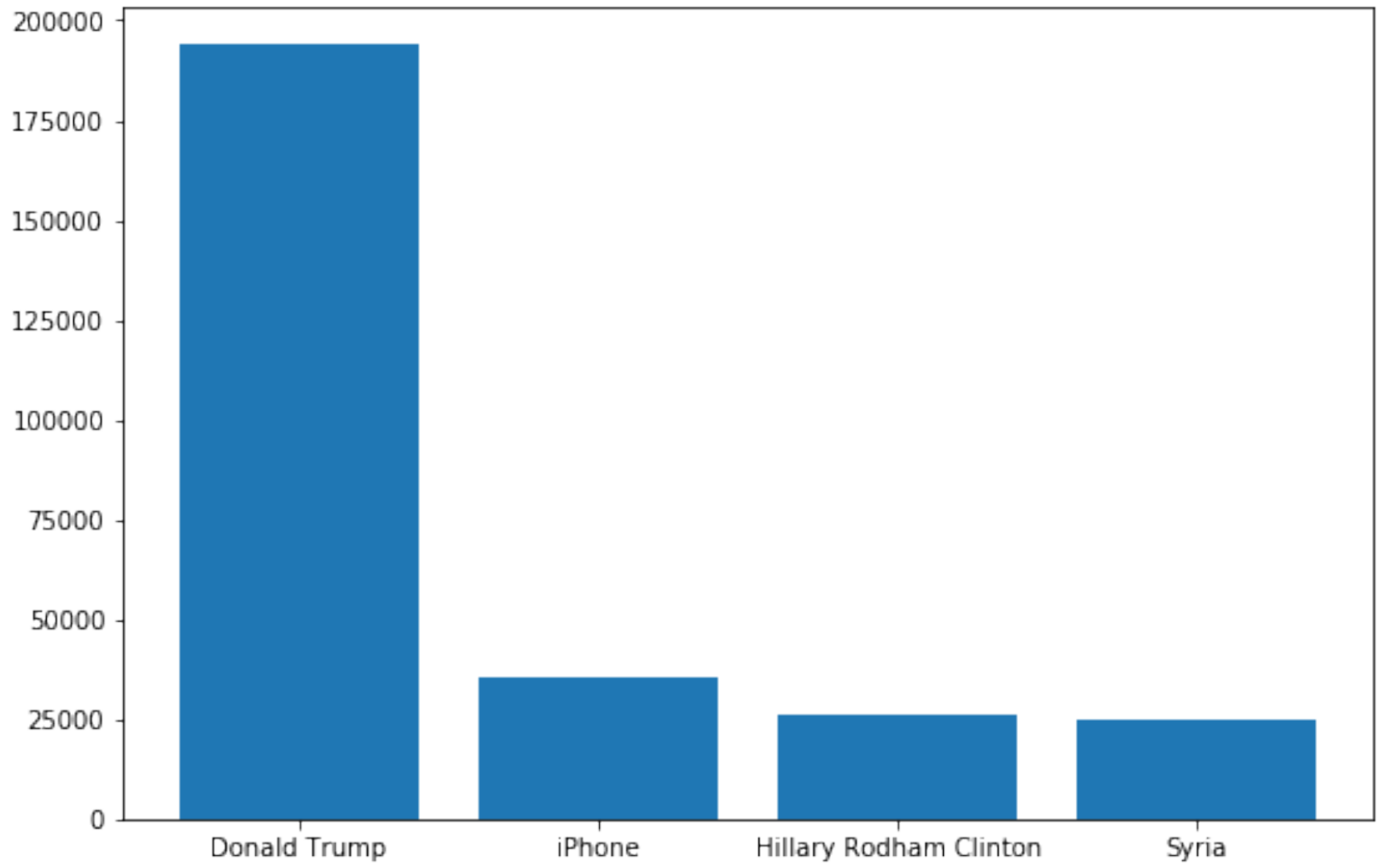

**FIGURE 2.** Distribution of the most 4 popular topics

The four most popular topics in this articles database are shown in Figure 2. According to this figure, most of the articles were about President Trump (roughly 190k articles), while the second most famous topic is "iPhone" (appears nearly 50k times). Tthe result showed that most of the articles in these databases are related to political (Donald Trump, Clinton, Syria). These topics are often categorized as "Top Stories" in the system. Moreover, after checking the time frame of the database, the oldest article was from October 2015 while the latest one was from November 2016, which was the last election campaign. Because of the reasons above, this dataset could be described as a collection of articles that heavily focus on political events, not only in the U.S. but also in the world.

## KAGGLE DATA CONNECTING PROCESS

Nine Kaggle datasets, excluded the news dataset, were connected by specific rules and conditions. As mentioned in the section above, the five datasets on crime in big cities were quite similar to one another, having only a few distinct expressions. With that convenience, the first step in the pre-processing procedure was joining these five datasets. The merged dataset contains the date, location, crime types and details, longitude/latitude, and original source where a sample came from. Meanwhile, some datasets, such as Global terrorism, have many attributes that were not common with the general crime record. This dataset has not only similar attributes as the merged states dataset but also a lot of attributes about penetrator, victim, weapon, intentions, and damages, making the joining process more complicated. Other datasets mentioned in Table II do not have this complexity since attributes on those datasets often focus on the same categories as states dataset, victim's information and penetrator's information. Based on the above conclusion, 27 attributes were chosen as attributes for the final dataset and these attributes were mentioned in Table IV.

**Table IV.** 27 attributes of the final crime dataset

| Attributes in final dataset 1 | | |
|---|---|---|
| Date | VIctimAge | PerpeAge |
| CrimeType | VictimRace | PerpeNationality |
| CrimeDetail | VictimGender | PerpeVicRelation |
| Lat | VictimDescription | Weapon |
| Long | TotalVictims | Motivation |
| LocDescription | PerpeMental | NewsCorverage |
| City | PerpeFlee | PropertyDamge |
| Street | PerpeRace | DataBase |
| State | PerpeGender | |

A merging procedure for each database was performed through these steps: First, attributes from the input database were compared with the attributes listed in table IV and identical attributes data were transferred, while missing attributes will be replaced with Null value. For the global terrorism dataset, only data related to an event in the U.S was chosen, leading to a very small number of data points related to terrorism. This procedure had successfully generated a collection of nearly 4.5 million samples and 27 attributes. Figure 3 is a pie plot that describes the distribution in the percentage of the final collection. As illustrated by the figure, crime reports datasets from Philly, San Francisco, Chicago account for nearly 75.0% of data points, while that number for the data from Denver is 10.9%, and the Homicide database is 15.0%. Moreover, the combination of terrorism dataset and police dataset only accounts for a very small percentage of the final dataset, which is around 1%. Because of that reason, it is quite difficult to notice those two datasets in the figure below

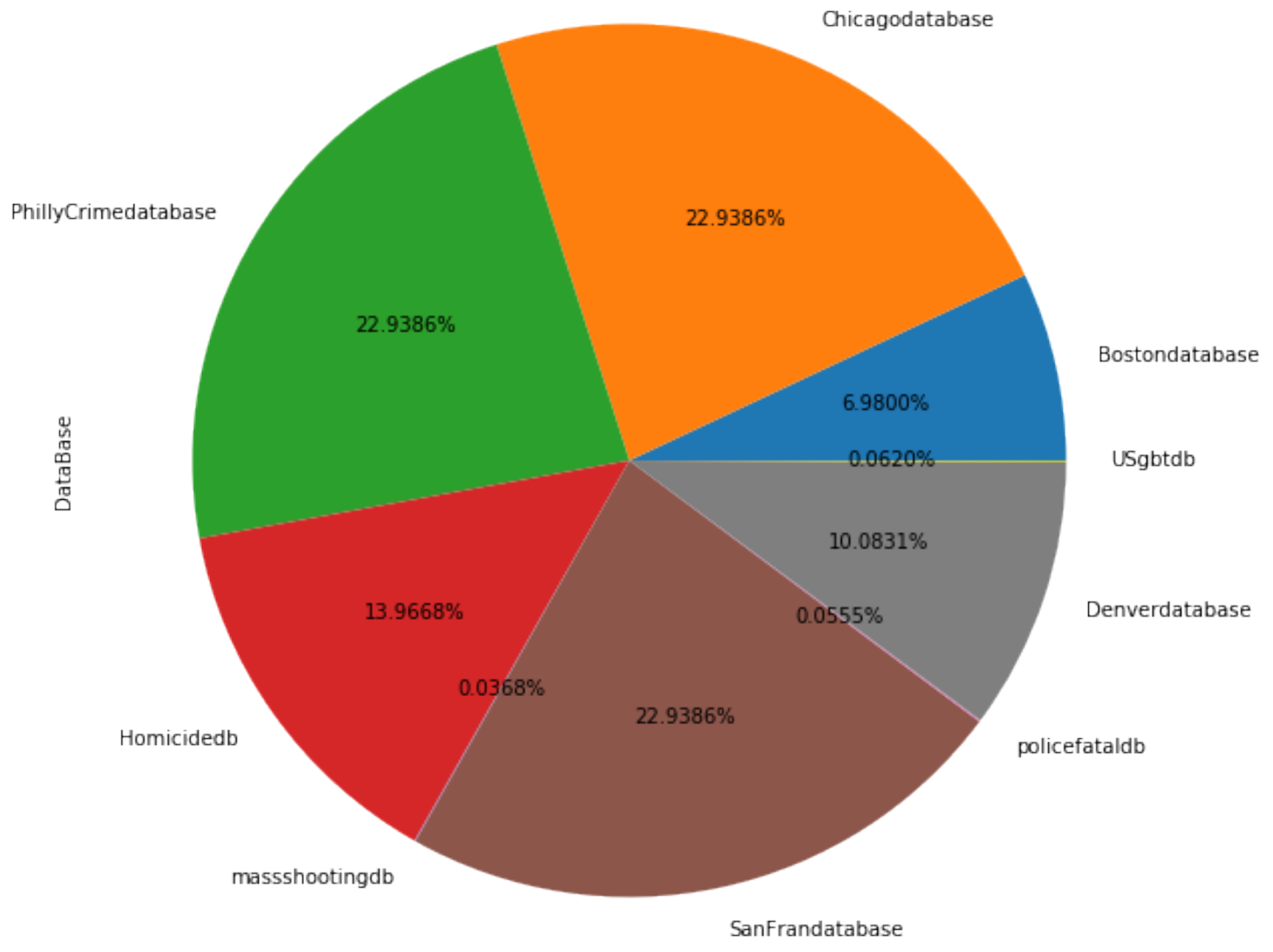

**Figure 3.** Distribution of the final crime dataset.

The next important step was sorting the value in "CrimeType" column into the corresponding category. Most of the crime types have several names, making clustering and creating a crime map a challenge. Auto theft, for example, can have several equivalent phases such as vehicle theft or motor vehicle theft; all elements in the set {larceny, theft, burglary, robbery} describe a robbery. This problem was solved by a four-step approach. The first step was to go over the most popular crime types, then organized similar values into one group, similar to how the above set contains words that mean robbery. In the second step, a program was used to find all data points in the dataset belong to a category, and subsequently marked those points with corresponding categories. However, some constraints were added to reduce the error of the mapping process, such as the one to help to separate "vehicle theft", "vehicle accident" and "accident" categories from each other. This procedure had successfully sorted crime reports into ordinary crime types, such as robbery, assault, or crime related to drugs. The result in Figure 4 shows that most of the cases in the dataset are related to assault, robbery, or drug. This fact is quite similar to the idea that assault and robbery were typical crime types in the society, where the main reason to commit a violation can be the "powerless" or the "powerful" of a human (RJ Michalowski,1985).

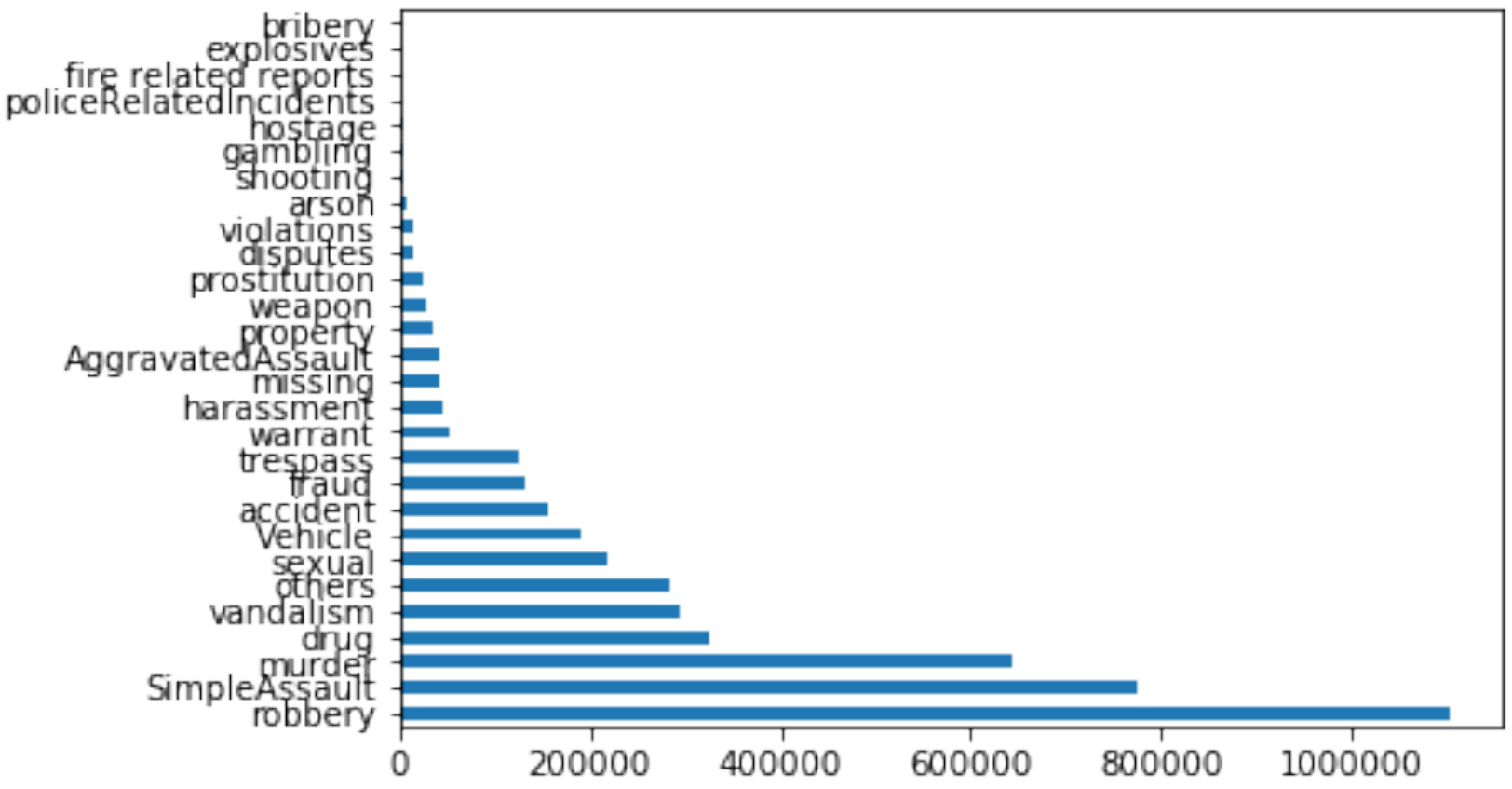

**Figure 4.** Crime categorizes distribution

Moreover, this mapping process was very useful to the processing News database procedure, especially in extracting articles related to crime among a collection of political, business, sport and entertainment news. The distribution of longitude and latitude in figure 5 also supports the fact that most of the data sample was located in the U.S, since longitude values ranged from -125 to -75, while latitude values were in the range of 30 to 50, for most of the times. Moreover, the centralized group is expected as the U.S shape since the team had preselected data only from U.S. However, there are a few mistakes, or outlier, could be observed in figure 5, and these errors should be corrected in future work. Lastly, most of the data are in the coastal areas or in the biggest cities, and it seems that crime density pattern follows population density patterns.

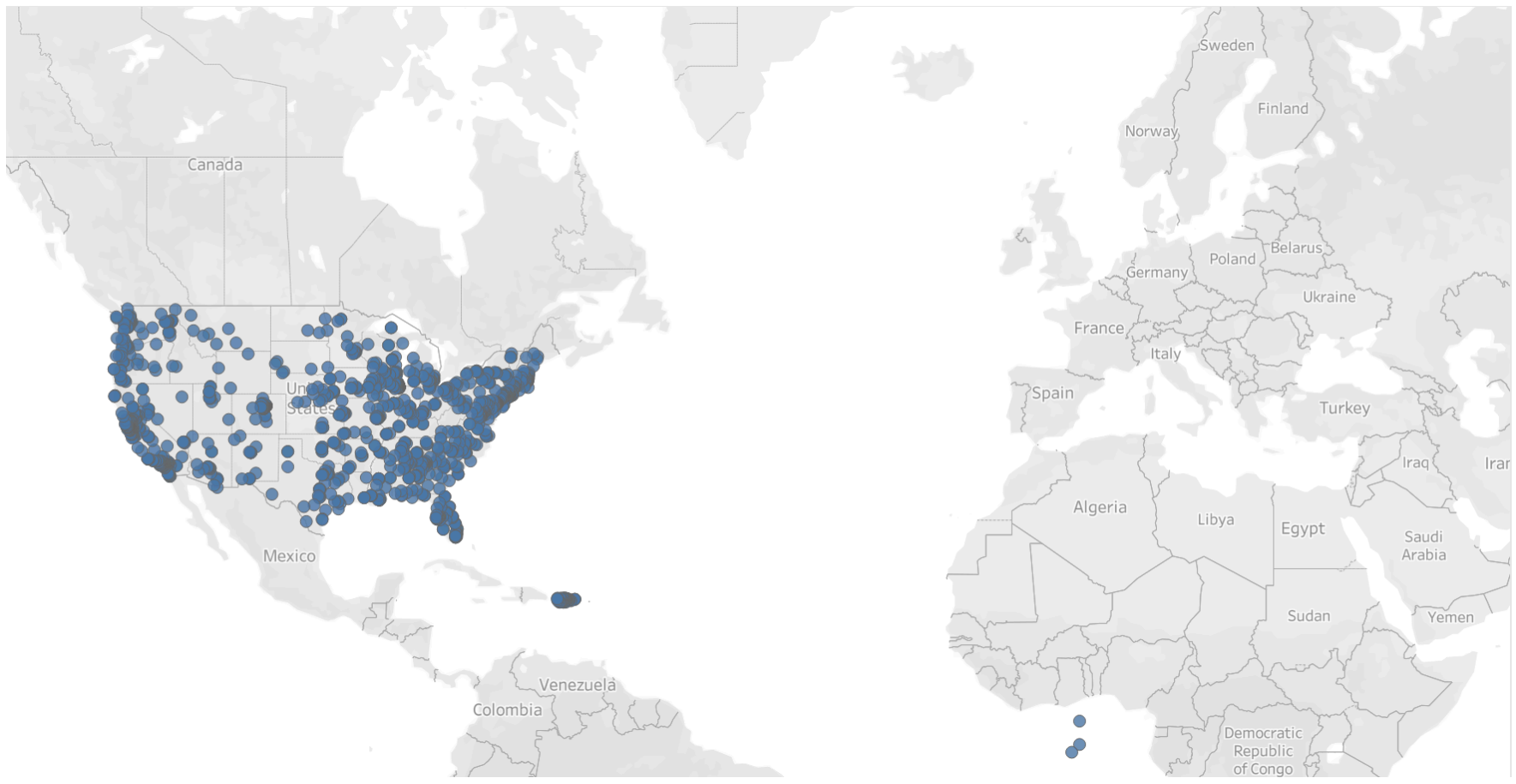

**Figure 5**. Longitude and Latitude distribution of recorded crime activities

## PROCESSING ARTICLES DATABASE

A similar merging procedure was used to merge the two articles dataset from Kaggle and from the Eager news projector. Data source, ID, Title, Publication, NewslineID, NewsOutledID, Author, Publishtime, OutletnName, Content, ArticleURL are attributes of the final dataset. When transforming an old data point into the new one in the final news dataset, missing attributes were replaced by Null value. The merged dataset could be divided into various topics, but only articles related to the crime were crucial to the research. With the goal of selecting only articles related to crime, a dictionary of 70 words was built after analyzing the most common keywords narrated to crime in the merged version of Kaggle datasets.

**Table V.** 70 words strongly related to crime.

| | Attributes in final dataset<br>1 | |
|---|---|---|
| robber | shot | prostitution |
| theft | fire | homicide |
| stol | bomb | manslaughter |
| steel | armed | assassination |
| thie | kill | damage |
| larcen | hijack | vandalism |
| burglar | fraud | criminal |
| embezzl | counterfeit | crime |
| weapon | forger | mischief |
| assault | battery | trafficking |
| arson | rape | kidnap |
| incident | offense | hostage |
| attack | forcible | barricade |
| victim | sex | ritual |
| terrorism | missing person | aggravated |
| police | runaway | dispute |
| perpetrat | accident | officer |
| shoot | vehicle | trespass |
| indecency | explosive | intimidate |
| disorder | bribe | harassment |
| violat | threat | |
| vargrancy | obsen | |
| loister | stalk | |

An article was categorized as a crime story when it had at least three words in the crime dictionary. On the other hand, there was some cross-over between legal drug activity and illegal drug activity, or between a vehicle accident and a vehicle theft, so an additional constraint was imported to eliminate articles about drug tests, business fights or accidents. Additionally, a word cloud was created after selecting articles related to crime and analyzing the distribution of words used in the that dataset, as shown in Figure 6. As in the figure, a lot of words are related to politic and violence. One big insight could be observed that there were many articles on President Trump and Islamic, a phenomenon that can be explained by the tension between the U.S. and Islamic countries. However, some results unrelated to crime can still be seen in the figure, such as “now”, “many” or “back”. Because of that reason, additional LDA analysis and clustering analysis were performed in next steps in order to clean the data more carefully and to reduce unwanted outliers.

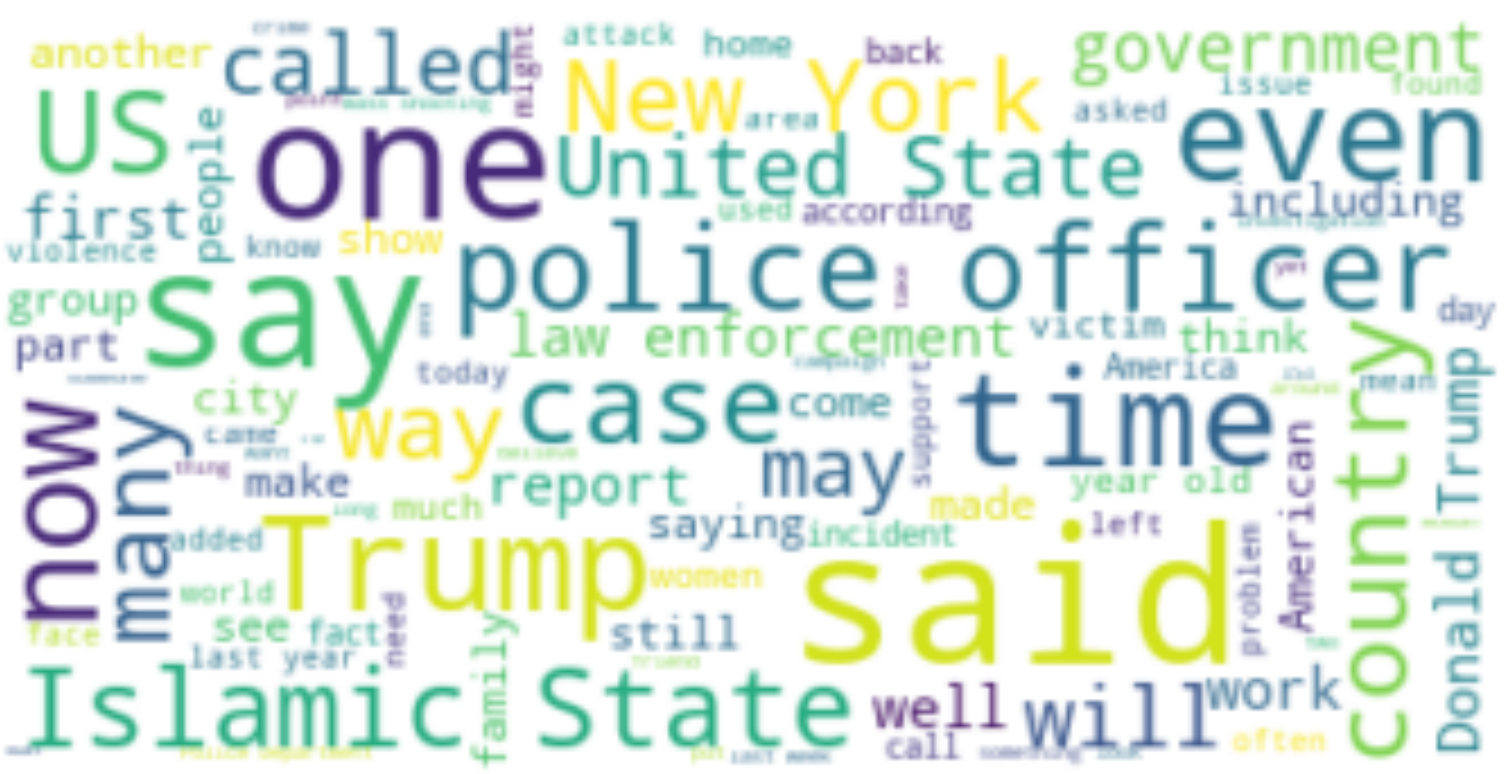

**FIGURE 6**. Popular words in the cleaned article dataset.

## TOPICS EXTRACTION WITH LDA

A methodology for extracting 30 topics from each article was built with “NLTK” and “Gensim” packages and this method had successfully extracted topics from 10929 articles in the final news dataset. The first step of this method was cleaning the content of every sample by removing unwanted characters such as space, punctuation, or comma. Secondly, articles with fewer than 1200 words were removed from the dataset, since the testing process on several articles with short lengths often returns a very small number of topics. Moreover, the content in each article was filtered by a function that removed number, stop words, and normalizing words. The Gensim’s models were used in the next step as the main tool to group related elements into one token and to explore sparse vectors. Moreover, the doc2bow() function was implemented to evaluate the occurrences of each word before transforming results into a sparse vector representing ids and occurrences of words since this function had been proven with good results in topic classification (J Novotny, 2017) . Finally, the LDA model was imported and executed to extract 30 topics per article. In table V, topics with the highest occurrence were listed, and the result in this table strongly is related to criminal activities.

**Table VI.** Statistics of Popular Topics extracted by LDA model

| Topic | Appearance in dataset |
|---|---|
| Attack | 7414 |
| Police | 5958 |
| Trump | 3575 |
| Black | 1913 |
| Officer | 1867 |
| Arrested | 1801 |

## ENTITIES EXTRACTION

An application was used to extract 30 entities from every single article in the news database. This application was designed in previous research, and only a small amount of modifications was implemented. In this algorithm, NLTK (Natural Language Toolkit) was used to help processing text data, removing numbers or special character, and normalizing words into standard form. In the next step, a list of tuples was generated, and in this list, every tuple contained a word and its associated part in the structure of the sentence. This collection of tuples was then used as the import for the entity extracting function. The entity extraction main function is a language model in English imported from the “SpaCy” package. This model helped a lot in applying the statistical system to assign labels to a group of tokens and also to give these topics numerical values (FNA AI Omran,2017). After been extracted, this entities list for articles was added to the database as new an attribute and was planned to be used on clustering and mapping progress.

## K-MEANS CLUSTERING ON NEWS COLLECTION

A clustering procedure for articles based on their contents and topics was designed and tested successfully with the use of TFIDF, K-Means, and DBSCAN algorithms. The Term Frequency-Inverse Document Frequency (TFIDF) method was used to create a matrix of documents based on the news collection. This algorithm determines the frequency of the word in the document and inverse frequency of the word in the whole corpus, then uses the measured value to rank the importance of each word relating to the document and vectorize the distribution of words in the article. The main goal of this function is to look for the most important words in each article, which are those with high frequency (TF) since words appeared frequently in all article were not helpful in distinguishing the articles. Due to that reason, the TF value was multiplied with the id f (inverse frequency of the word in the whole corpus) to get the final TFIDF score. For example, the word “drugs” appeared often in drugs related articles, thus getting a high TF score and a high IDF score since the word did not appear often in the rest of the corpus. On the other hand, the word “the” had a low TDIFD score since it appeared very often in articles, leading to high TF score for each article but low IDF score in the whole corpus. When adjusted for the parameter of the TFIDF function, the “min_df” value is 5, meaning the terms with smaller frequency will be removed. On the other hand, terms that appear in more than 95% of documents would be removed and only 60 features were chosen to be the attributes of the vectorization of articles’ contents, as in table VII.

**Table VII.** 60 features for vectorizing procedure.

| | Features | |
|---|---|---|
| Abddeslam | Ago | Arrest |
| According | Air | Arrested |
| Accused | Airport | Assault |
| Act | Al | Attack |
| Action | Allegedly | Attorney |
| Activist | America | Authority |
| Afp | Announced | Black |
| Afternoon | Ap | Campaign |
| Agency | Area | Case |
| | Armed | Chicago |
| Child | Day | Obama |
| City | Death | Officer |
| Clinton | Family | People |
| Country | Group | Police |
| Court | Gun | President |
| Crime | Killed | Said |
| Dallas | Mateen | Say |
| Time | Muslim | Shooting |
| Trump | New | State |
| Woman | Year | |

K-means algorithm, with the ability to slice the dataset into k number of distinct subgroups based on the distance between samples and the centroid of the cluster, was used to group similar articles together, based on the features(words) vector from TFIDF function. These distances were used to determine the similarity between data points. The closer the distance, the more similarity between two data points. Due to that reason, different numbers of clusters were tested to find the most suitable number to fit the distribution of the dataset. In Figure 6, it was not very clear elbow how the slope could change from very big to very small. However, there were several points where the slopes had significant changes, such as at 64, a significant point where the SSE value started lowering.

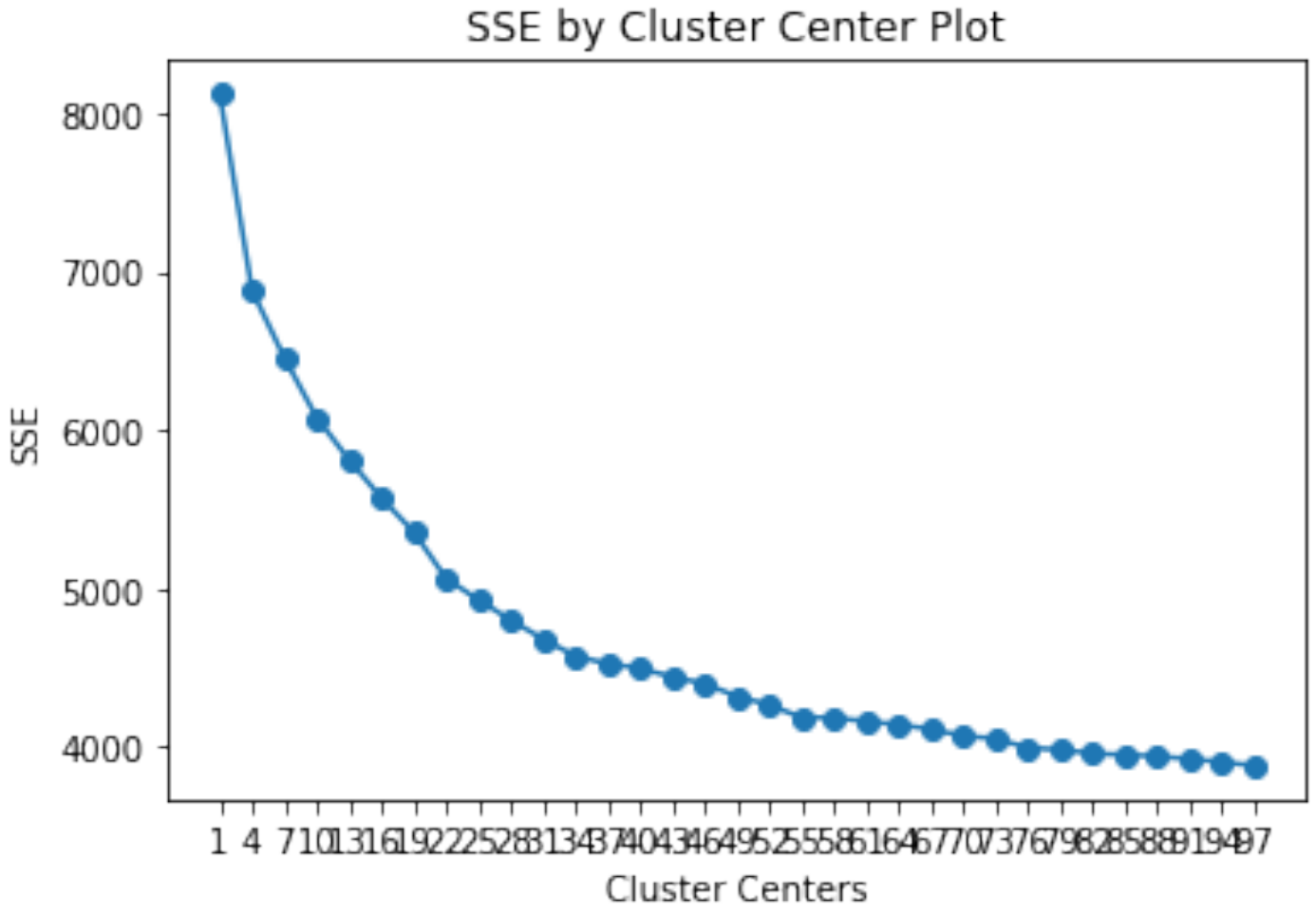

**FIGURE 6**: Relationship betwenn cluster centers and SSE

Density-based spatial cluster of applications with noise (DBSCAN) was also used to cluster the dataset. Different from the K-Means algorithm that focused on finding a number of nearest neighbors, the DBSCAN focused on grouping data to one cluster based on the distance from the center. DBSCAN function will group together points that are close to each other based on their distance measurement and a minimum number of points in a density, and two close points, or two close cluster, could be grouped together into new clusters. Moreover, this algorithm can help pointing out outliers, which lie alone in low density regions Despite the fact that DBSCAN was not the best fit for topic clustering, DBSCAN had successfully cluster the datasets into several main groups.

## RESULTS AND CONCLUSION

The research had successfully combined multiple datasets on Kaggle related to criminal activity, as illustrated in Table I and II, into a new dataset combined all information from nine original collections, without losing any significant information. Moreover, similar expressions of a single crime in different datasets were also recognized, and these expressions had been mapped into meaningful new values, which called Crime Map. Based on the combined dataset, simple assault is the most common criminal type, followed by robbery and drug. This Crime Map was used to select articles related to crime in the news database. The result was very promising since Figure 6 clearly shows that selected articles were mainly about criminal activities.

Two methods used to extract topics and entities of articles were tested and these methods returned acceptable results. Using the extracted topics values from the LDA model, two cluster plots of the dataset were created by TFIDF and K-Means algorithms. Each plot conveys the relationship of 50 clusters represented in 2D dimension. In Figure 7, the result of the TSNE cluster plot is quite good, since there are several separated clusters that can be found on the plot. On the other hand, the PCA plot had also expressed distant clusters and several outliners. However, by scrutinizing on the middle of the PCA plot, a set of mixed values from different clusters can be observed. This can suggest that the original data had a high number of dimensions. As a result, when the data was illustrated in 2-D dimension, several datapoints could be very close one another (H Xu, 2010). Another explanation for this pattern was in the middle of the plot, where the data points were actually similar to one another, since some topics could have several similar characteristics.

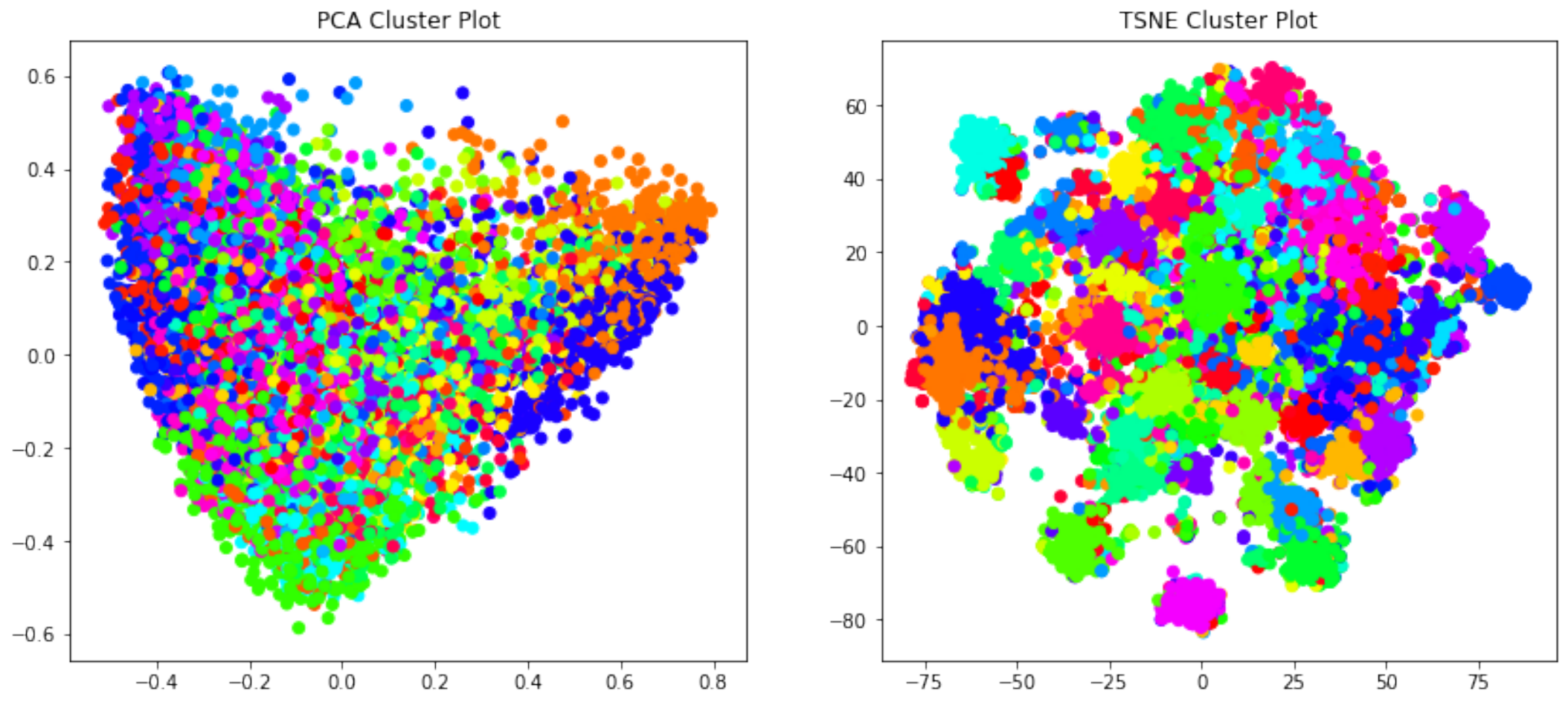

**FIGURE 7**. Clustering results by K-means with 64 clusters and 60 features

The clustering results from K-means algorithms contained keywords from each cluster, and these keywords were used to measure the quality of clustering models. For example, here is the list of keywords in clusters 5, 37, and 41:

*Cluster 5: day, trump, American, country, time, people, president, year, said, new.*
*Cluster 37: new, year, assault, court, according, people, police, said.*
*Cluster 41: people, gun, black, officer, year, said, shooting, city, police, Chicago.*

From the three examples above, cluster 5 was an outlier cluster, where the article has not really focused on criminal activities. Cluster 37 can be named as a cluster of assault records, which could have been processed by the court since the word "court" appears as a keyword of the cluster. Moreover, cluster 41 could have grouped articles related to criminal activities related to black people in Chicago with a gun involved in the crime screen since "gun", "black" or "Chicago" are popular keywords in the cluster dictionary.

DBSCAN with "eps" value (the distance from the point in cluster in which algorithm looks to find other points that are still not in the cluster to be added to) equal to 1 and min samples value equals to 10 had also used to cluster the topics of news collection ,and the algorithm had returned quite similar results compared to K-means'. The DBSCAN clustering model had returned 90 clusters, with 6552 points belongs to ID -1- outliers, followed by the biggest cluster with 407 points and the second with 388 points. On the other hand, there were several clusters with only 10 or 9 points. These small clusters could be suspected as outliers since they only contained a very small number of articles that may not related to any criminal activities at all, but in many cases those clusters would be the legit ones.

These extracted attributes had been used to divide the whole dataset into separated context-related clusters, and the output clusters had been examined to help to prove the efficiency of the extracting methods. By checking the keywords of both central cluster and outlier cluster, the team can conclude that these clusters strongly grouped related articles together, and the more centralized a cluster was, the more likely it could be related to criminal activities.

In the future, those clustered groups of articles could be used to assign crime type to new articles or criminal records or to connect new datasets with the combined Kaggle dataset. For example, crime type would be given based on clusters' topics, while latitude and longitude can be collected from locations extracted among entities, or the text could be the description of the crime. The big dataset, after being joined, could be used to analyze crimes activity or be used to connect crime records into a crime map.

## ACKNOWLEDGEMENT

This research was supported in part by the NSF grant IIS-8142183.